# CHEMICAL CAPPING SYNTHESIS OF NICKEL OXIDE NANOPARTICLES AND THEIR CHARACTERIZATIONS STUDIES

**M.Nowsath rifaya, T.Theivasanthi\* and M.Alagar**

*Center for Research and Post Graduate Department of Physics, Ayya Nadar Janaki Ammal College, Sivakasi - 626124, Tamilnadu, India.*

**ABSTRACT:** This work reports aspect related to chemical capping synthesis of nano-sized particles of nickel oxide. It is a simple, novel and cost effective method. The average particle size, specific surface area, crystallinity index are estimated from XRD analysis. The structural, functional groups and optical characters are analyzed with using of SEM, FTIR and UV- visible techniques. XRD studies confirm the presence of high degree of crystallinity nature of nickel oxide nanoparticles. Their particle size is found to be 12 nm and specific surface area (SSA) is $74 m^2 g^{-1}$. The optical band gap energy value 3.83ev has also been determined from UV-vis spectrum.

## 1. Introduction

Since the innovation of nanochemistry in past decades, numbers of materials in nano-scale have been synthesized via many methods [1]. Nanomaterials have been widely used in various fields, such as photoelectric, recording materials, catalysts, sensors, ceramic materials, etc., due to their special structures and properties [2–5]. In particular, nanosized nickel oxides exhibit particular catalytic [6, 7], anomalous electronic [8, 9] and magnetic [10, 11] properties. Another important application of NiO is in battery systems [12, 13]. Non-stoichiometric nickel oxide is a good P-type semiconductor owing to its defect structure [14], a potential gas sensor for $H_2$ [15] and high active in the degradation of phenol and phenolic derivatives [16]. These applications can be enhanced by decreasing the particle size (preferably to less than 20 nm) and are highly dependent on particle size; the precise control of the size and distribution in a nanometer region is required. Recently, several methods have been developed to prepare ultrafine nickel oxide powder, including low-pressure spray pyrolysis [17], surfactant-mediated method [18], simple liquid phase process [19] and other techniques [20–22] .This work reports the synthesis of nickel oxide nanoparticles by chemical capping method using nickel chloride, alanine, ethanol and ammonia. Using alanine – an aminoacid in synthesis process is the novelty of this work and the end product nickel oxide nanoparticles will be bio-compatible in nature and will be very useful for bio-sensor related applications.

---

**\*Corresponding author.** *E-mail*: theivasanthi@pacrpoly.org  Cellphone: +91-9245175532

## 2. Experimental Details

The NiO nanoparticles are synthesized by chemical capping method. In the chemical capping technique, capping molecules terminate the nano particle growth by blocking the active sites. For the synthesis of NiO nano particles, alanine was used as the capping molecule. Two master solutions were prepared: solution A, consisting of 0.1M solutions of $NiCl_3$ in ethanol and solution B consisting of 0.001M solution of alanine in ammonia. Solution A was added drop wise to solution B with continuous stirring and the mixture was maintained at a temperature of 80°C at atmospheric pressure for 20 hrs. Due to the insolubility of synthesised nano particles in ethanol and ammonia, the green particles settled down. The nano particles were collected and washed several times in high purity water to remove the excess ions ($NH^+$ and $CL^-$). The samples were dried at 60°C in air to remove moisture contents. The samples were prepared by heating this $Ni(OH)_3$ nanoparticles at 350°C for 90min.

X-ray diffraction (XRD) measurements were performed using a MAC Science MXP18 diffractometer with Cu Kα radiation ($\lambda$ = 1.5405Å) at 40 kV and 30mA with a scanning speed in $2\theta$ of $4°min^{-1}$. The crystallite sizes of nickel oxide were estimated using the Scherrer equation. The surface morphology of nickel oxide nanoparticles were observed by means of a scanning electron microscope (JSM-6330TF) operated at 10 kV. Fourier transform infrared spectra of green nickel oxide nano particles were recorded using

SHIMADZU spectro photometer with KBr pellet technique from 4000cm$^{-1}$ – 400cm$^{-1}$. For U-V absorption spectra, these samples are dissolved in glycerin with the help of magnetic stirrer for an half an hour at room temperature and then it was examined by UV-vis spectrophotometer.

## 3. Results and discussions
### 3.1. X-Ray Diffraction Studies

The grown nanoparticles were characterized by powder X-ray diffractometer. Fig 1 shows the X-ray diffraction spectrum of NiO sample. This shows crystalline structure with 5 peaks. The XRD pattern shows a significant amount of line broadening which is a characteristic of nanoparticles. The XRD pattern exhibits prominent peaks at 37.3°, 43.3°, and 63°. The crystal size can be calculated according to *Debye-Scherrer formula* [23].

$$D = \frac{0.9\,\lambda}{\beta \cos\theta} \quad \ldots\ldots\ldots\ldots\ldots\ldots\ldots\ldots\ldots\ldots\ldots \text{(1)}$$

Where k=0.89, λ is the wavelength of the Cu-Kα radiations, β is the full width at half maximum and θ is the angle obtained from 2θ values corresponding to maximum intensity peak in XRD pattern. The mean crystal size of NiO nanoparticle is 12.2 nm.

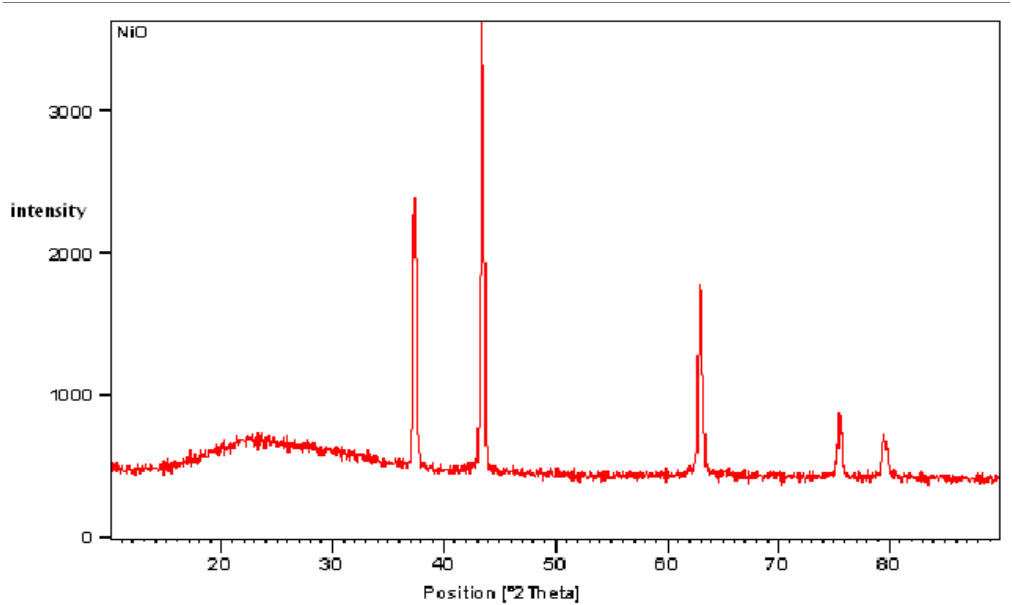

Fig. 1. XRD pattern of NiO Sample.

The inter planar distance was calculated using *Bragg's Law* [24].

$$2d \sin \theta = n\lambda \quad \quad (2)$$

Where n was taken as 1. The value of d for the most intense peak was 2.08432Å. The diffraction peaks thus obtained from X-ray diffraction data are in good agreement with the standard pattern of NiO [1].

### 3.2. XRD - Crystallinity Index

It is generally agreed that the peak breadth of a specific phase of material is directly proportional to the mean crystallite size of that material. From our XRD data, a peak broadening of the nanoparticles is noticed. The average particle size, as determined using the Scherrer equation, is calculated to be 12.2 nm. Crystallinity is evaluated through comparison of crystallite size as ascertained by SEM particle size determination. Crystallinity index Eq. is presented below:

$$I_{cry} = \frac{D_p(SEM, TEM)}{D_{cry}(XRD)} (I_{cry} \geq 1.00) \quad \quad (3)$$

Where $I_{cry}$ is the crystallinity index; $D_p$ is the particle size (obtained from either TEM or SEM morphological analysis); $D_{cry}$ is the particle size (calculated from the Scherrer equation).

Table.1: The crystallinity index of Nickel Oxide Nanoparticles

| Sample | Dp (nm) | Dcry (nm) | Icry (unitless) | Particle Type |
|---|---|---|---|---|
| NiO Nanoparticles | 97 | 12 | ~8.08 | Polycrystalline |

Table.1. displays the crystallinity index of the sample that scored higher than 1.0. The data indicate that the NiO is highly crystalline. If $I_{cry}$ value is close to 1, then it is assumed that the crystallite size represents monocrystalline whereas a polycrystalline have a much larger crystallinity index [25].

### 3.3. XRD - Specific Surface Area

Specific surface area (SSA) is a material property. It is a derived scientific value that can be used to determine the type and properties of a material. It has a particular importance in case of adsorption, heterogeneous catalysis and reactions on surfaces. SSA is the Surface Area (SA) per mass.

$$SSA = \frac{SA_{part}}{Vpart * density} \quad \ldots\ldots\ldots\ldots\ldots\ldots\ldots\ldots\ldots\ldots (4)$$

Here Vpart is particle volume and SApart is particle SA [26].

$$S = 6 * 10^3 / D_p \rho \quad \ldots\ldots\ldots\ldots\ldots\ldots\ldots\ldots\ldots\ldots (5)$$

Where S is the specific surface area, Dp is the size of the particles, and $\rho$ is the density of NiO 6.67 g cm$^{-3}$ [27]. Mathematically, SSA can be calculated using these formulas 4 and 5. Both of these formulas yield same result. Calculated value of (SA = 468 nm$^2$, Volume = 951 nm$^3$ and SSA = 74 m$^2$ g$^{-1}$) prepared NiO nanoparticles are presented in Table.2.

Table 2. Specific Surface Area of Nickel oxide Nanoparticles

| Particle Size (nm) | Surface Area (nm$^2$) | Volume (nm$^3$) | Density (g cm$^{-3}$) | SSA (m$^2$g$^{-1}$) | SA to Volume Ratio |
|---|---|---|---|---|---|
| 12 | 468 | 951 | 6.67 | 74 | 0.49 |

### 3.4. SEM Analyses

The surface morphological features of synthesized nanoparticles were studied by scanning electron microscope. Fig 2 shows the SEM image of NiO nanoparticles with magnification of 5000. The instrumental parameters, accelerating voltage, spot size, and magnification and working distances are indicated on SEM image. The results indicate that mono-dispersive and highly crystalline NiO nanoparticles are obtained. The appearance of some particles is in spherical shape and some are in rod shape. We can observe that the particles are highly agglomerated and they are essentially cluster of nanoparticles. The SEM picture indicates the size of polycrystalline particles. The observation of some larger nanoparticles may be attributed to the fact that NiO nanoparticles have the tendency to agglomerate due to their high surface energy and high surface tension of the ultrafine nanoparticles. The fine particle size results in a large surface area that in turn, enhances the nanoparticles catalytic activity. So we can conclude that the prepared NiO particles are in nanometer range. The average diameter of the particle observed from SEM analysis is 97 nm, which is larger than the diameter predicted from X-Ray broadening.

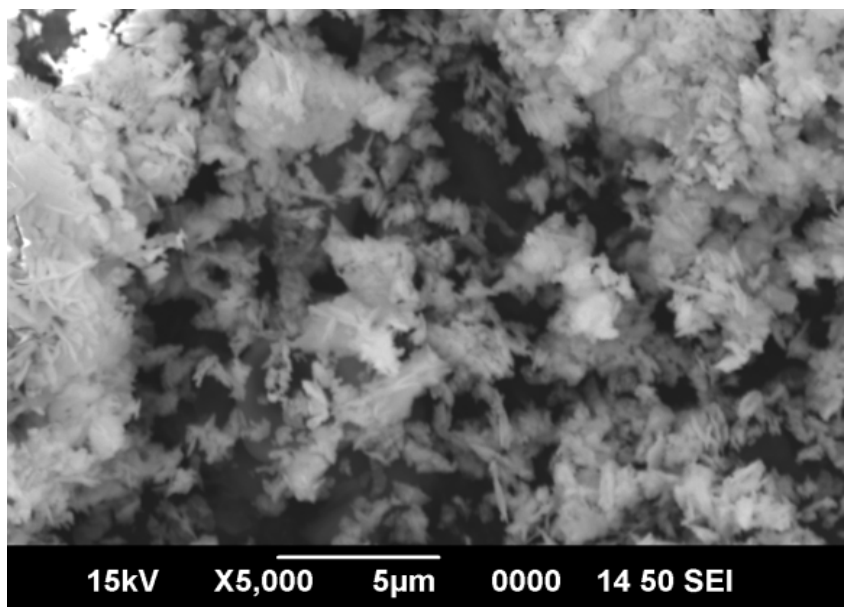

Fig. 2. SEM image of nickel oxide nanoparticles: 5000 magnification.

## 3.5. FTIR Analyses of Nickel oxide nanoparticles

Fig 3 shows Fourier transformed spectrum of NiO nano particles at room temp. The spectrum was recorded in the range of 4000cm$^{-1}$ – 400cm$^{-1}$. The FTIR spectrum shows the characteristics peaks at 418.57cm$^{-1}$, 677.04 cm$^{-1}$, 1629.90 cm$^{-1}$, 2924.18 cm$^{-1}$,

3431.48 cm$^{-1}$. The band at 418.57 cm$^{-1}$ reveals the presence of NiO. Despite drying this sample contained traces of water (peaks at 1629.90 cm$^{-1}$, 3431.48 cm$^{-1}$). There is no peak indicating the presence of ammonia and chloride as the sample was washed with high purity water. This shows that the sample contains no impurity. The same chemical composition is confirmed by the XRD pattern.

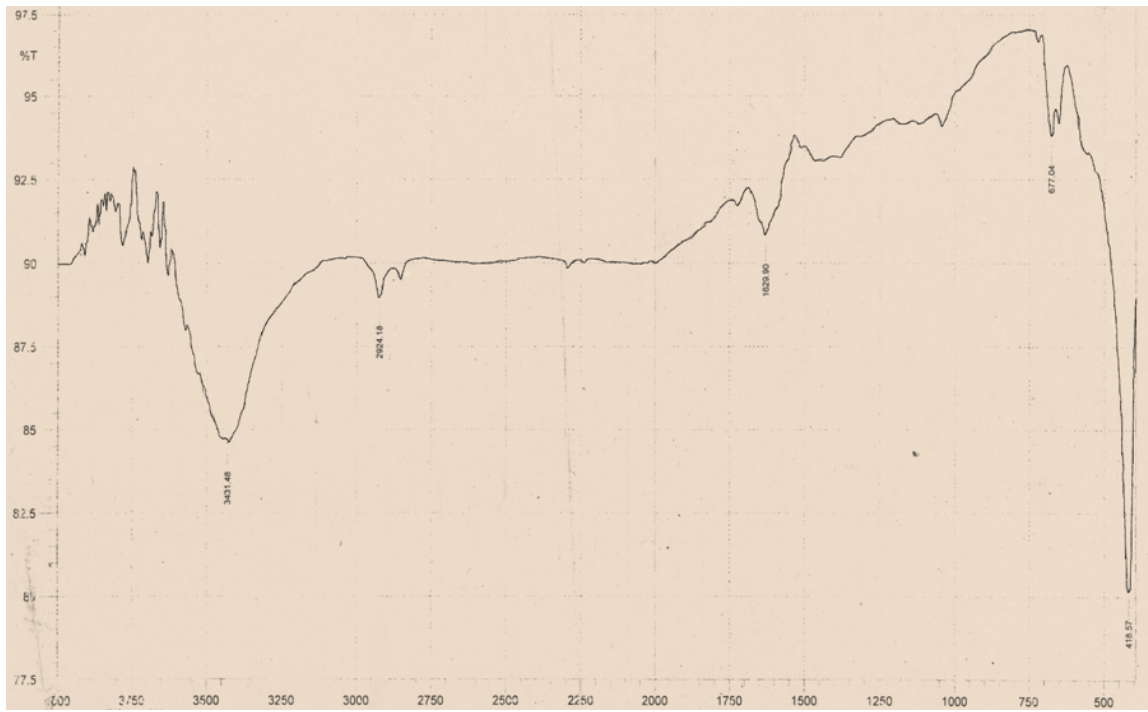

Fig 3. FTIR spectrum of the NiO nanoparticle

### 3.6. UV-Vis Studies of Nickel oxide nanoparticles

Once nano nickel oxide is confirmed then the optical characterization is important. The optical absorption spectrum of nickel oxide is shown in fig 4. It can be seen that the strongest absorption peak of the as-prepared sample appears at around 342 nm, which is fairly blue shifted from the absorption edge of bulk NiO nanoparticle. The bandgap energy calculated from UV-absorption is 3.83ev. This value is higher than bulk NiO i.e. $E_{gb}$=3.74ev. So it is highly agreed that the synthesized NiO is in nano scale.

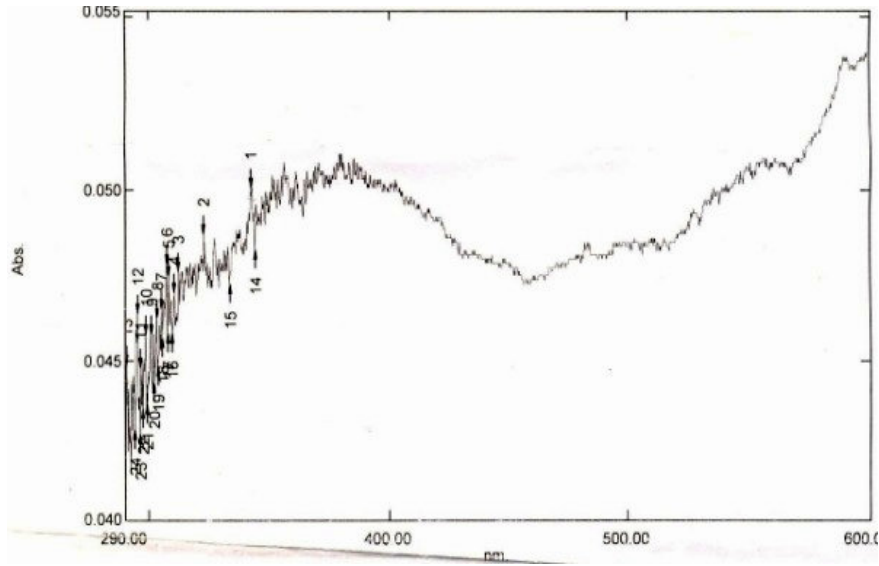

Fig 4. UV-Vis absorbance spectrum

Energy gap value was calculated using Tauc relation [28]

$$\alpha h\nu = A(h\nu - E_g)^n \quad \text{................................................. (6)}$$

where A is the parameter that depend on the interband transition probability, $E_g$ is he optical band gap, h is the incident light and n is the number characterizing the nature of the transition process; n=2 for direct transition and n=1/2 for indirect transition. The optical absorption spectrum provides information about concerning the size. Fig 5 shows the optical band gap. So from the estimated optical band gap of the nano NiO and the shift in the band gap between the bulk and the nanoparticles, we can calculate the size of the nanoparticle as given in the following relation,

$$E_{gn} = [E_{gb}^2 + \{2h^2 E_{gb}/R^2\}m^*]^{1/2} \quad \text{........................ (7)}$$

where R is the radius of the quantum size particles i.e. nanoparticles, $E_{gn}$, $E_{gb}$ are the band gap of nano 3.83ev and bulk 3.74ev system respectively and m* be the effective mass. Using these values, the average size has been calculated as 48nm. When the band gap energy increases, the particle size decreases. From this we can conclude that the synthesized particle is a nanoparticle.

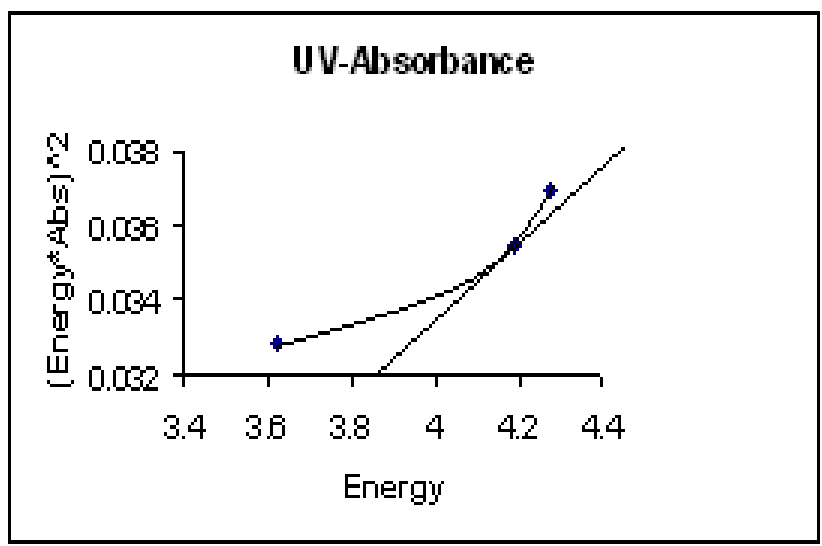

Fig 5. UV-Vis Band gap Energy

## 4. Conclusions

Nanostructured particles of NiO have been successfully synthesized through chemical capping method using nickel chloride, alanine, ethanol and ammonia. Alanine has served as capping molecule and blocked the active sites at growing surfaces. XRD, SEM, FTIR, UV-Vis characterizations studies have also been done for the synthesized nanoparticles. XRD results estimate the average particle size as 12 nm and the specific surface area as 74 $m^2$ $g^{-1}$. It also confirms the high degree crystallinity nature of the prepared sample. SEM confirms that the particles are in nano size and the appearance of some particles are in spherical shape and some are in rod shape. The FTIR spectrum confirms the presence of NiO nanoparticles. Energy band gap value 3.83ev has been calculated from UV-absorption. This simple, novel and cost effective chemical capping synthesis method will be useful for industries for the preparation of nickel oxide nano-sized particles.

## Acknowledgements

The authors express immense thanks to staff & management of *PACR Polytechnic College*, Rajapalayam, India and *Ayya Nadar Janaki Ammal College*, Sivakasi, India for their valuable suggestions and assistances.